\documentclass[lettersize,journal]{IEEEtran}
\usepackage{amsmath,amsfonts}
\usepackage{algorithmic}
\usepackage{algorithm}
\usepackage{array}
\usepackage[caption=false,font=normalsize,labelfont=sf,textfont=sf]{subfig}
\usepackage{textcomp}
\usepackage{stfloats}
\usepackage{url}
\usepackage{verbatim}
\usepackage{graphicx}
\usepackage{xcolor}
\usepackage{multirow}
\usepackage{url}
\usepackage{cite}
\usepackage{flushend}
\begin{document}

\title{The Weight of a Bit: EMFI Sensitivity Analysis of Embedded Deep Learning Models}

\author{Jakub Breier,
\thanks{J. Breier is with TTControl GmbH, Vienna, Austria.
E-mail: jbreier@jbreier.com.}
Štefan Kučerák,
Xiaolu Hou
\thanks{Š. Kučerák and X. Hou are with the Faculty of Informatics and Information Technologies, Slovak University of Technology, Slovakia.
X. Hou is also affiliated with the State Key Laboratory of Blockchain and Data Security , Zhejiang University.
E-mail: xkucerak@stuba.sk, houxiaolu.email@gmail.com.}
\thanks{This work was supported by the Open Research Fund of The State Key Laboratory of Blockchain and Data Security, Zhejiang University under grant no. A2566.}}

\markboth{Journal of \LaTeX\ Class Files,~Vol.~14, No.~8, August~2021}%
{Shell \MakeLowercase{\textit{et al.}}: A Sample Article Using IEEEtran.cls for IEEE Journals}


\maketitle

\begin{abstract}
Fault injection attacks on embedded neural network models have been shown as a potent threat.
Numerous works studied resilience of models from various points of view.
As of now, there is no comprehensive study that would evaluate the influence of number representations used for model parameters against electromagnetic fault injection (EMFI) attacks.

In this paper, we investigate how four different number representations influence the success of an EMFI attack on embedded neural network models.
We chose two common floating-point representations (32-bit, and 16-bit), and two integer representations (8-bit, and 4-bit).
We deployed four common image classifiers, ResNet-18, ResNet-34, ResNet-50, and VGG-11, on an embedded memory chip, and utilized a low-cost EMFI platform to trigger faults.
Beyond accuracy evaluation, we characterize the injected fault pattern by analyzing the bit error rate, the spatial distribution of corrupted bytes, and the prevalence of \texttt{0xFE}/\texttt{0xFF} byte values across formats, identifying the mechanisms responsible for the observed differences in resilience.
Our results show that while floating-point representations exhibit almost a complete degradation in accuracy (Top-1 and Top-5) after a single fault injection, integer representations offer better resistance overall.
In particular, the 8-bit representation on a relatively large network (VGG-11) retains Top-1 accuracy of around 70\% and Top-5 at around 90\%. 
\end{abstract}




\section{Introduction}
\IEEEPARstart{E}{mbedded} implementations of neural networks are gaining popularity with frameworks such as TinyML~\cite{lin2023tiny}, allowing a real-time, efficient execution in the Edge.
The applications range from the Internet of Things to mobile devices to autonomous systems, offering significant benefits: reduction in latency, improved data privacy, removing the necessity of a constant network connection, to name a few.
These lightweight frameworks take pre-trained models and optimize them with various techniques, such as quantization and pruning~\cite{liang2021pruning}, to fit in the resource-constrained hardware.
The result is the ability of small microcontrollers with only a few kilobytes of RAM to run models for image classification, anomaly detection, and others, effectively moving away from the classical cloud setup.

This naturally introduces new security and reliability challenges due to malicious and environmental influences stemming from the physical accessibility of these devices.
Fault injection attacks (FIAs) are one notable threat where the attacker disturbs the operation of the device to introduce errors, either in the data or execution flow~\cite{breier2022practical}.
Electromagnetic fault injection (EMFI)~\cite{dumont2019electromagnetic} is a fault injection technique that is non-invasive and does not need a sophisticated, expensive device to perform.
The attacker uses a high-power pulse generator to inject a sharp electromagnetic pulse, introducing transient faults in the device's electronic circuits. 
These faults can either cause bit flips in the memory or skip the instructions being executed, resulting in the misbehavior of the neural network model, as shown in~\cite{goswami2025emfi}.

When deploying neural networks on a microcontroller, one needs to consider the numerical representation of model parameters.
To achieve memory and power savings, it is standard to use reduced-precision formats: for instance, 32-bit floating-point weights can be reduced to 16-bit floats, 8-bit integers, and in extreme cases even to 4-bit or binary weights~\cite{hashemi2017understanding}.
While such precision reduction/quantization introduces an approximation error, when used properly, a quantized model can retain almost the same accuracy as the original one~\cite{nagel2021white}.
At the same time, quantization changes the fault tolerance characteristics of a network, leading to different vulnerability profiles under fault injection~\cite{guilleme2025sfi4nn}.
Intuitively, a bit flip in an integer weight has a bounded effect, especially since the weight range is limited by the quantization scale, whereas in a floating-point weight, a bit flip in the exponent bits can have a significant influence on the network's output, as was investigated in~\cite{hong2019terminal}.
In integer formats, the most significant bits are the critical ones for a network’s computation, while in floating-point, the exponent and higher-order mantissa bits carry the most significance. 
Furthermore, larger models provide inherent redundancy that compensate for some faults.

The open question, and the main focus of this paper, is how these factors play out in practice: 

\vspace{0.2cm}
\textit{To what extent does the choice of weight number format (from high-precision to aggressively quantized) influence a neural network’s resilience against EMFI attacks?}

\vspace{0.2cm}
Despite growing awareness of both quantization techniques and fault injection threats, this specific question remains only partially answered by prior research. 
Most existing studies of neural network fault tolerance have considered a single representation (often full 32-bit precision, or a very low-bit scheme in isolation), rather than comparing formats, thus a systematic investigation is warranted~\cite{guilleme2025sfi4nn,goswami2025emfi}.

\begin{figure*}[b!]
    \centering
    \includegraphics[width=0.95\linewidth]{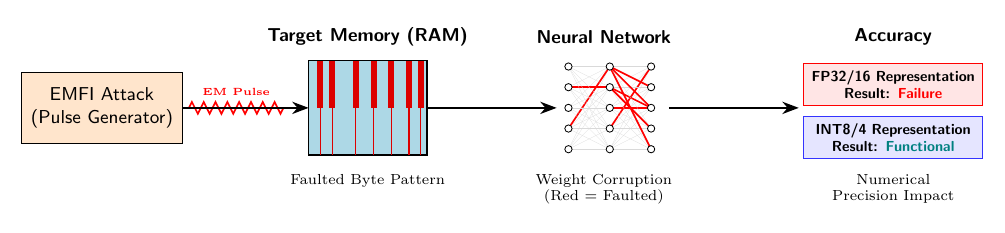}
    \caption{Overview of the EMFI analysis done in this paper.}
    \label{fig:method_overview}
\end{figure*}

\textbf{Our contribution.} In this work, we present an extensive empirical study of how weight representation influences the resilience of neural networks to electromagnetic fault injection.
We consider four widely used numerical formats for model weights: 32-bit floating point (FP32), 16-bit floating point (FP16), 8-bit integer (INT8), and 4-bit integer (INT4), spanning the range from full precision to aggressive quantization.
We evaluate four representative convolutional neural network (CNN) architectures---ResNet-18, ResNet-34, ResNet-50~\cite{he2016deep}, and VGG-11~\cite{simonyan2014very}---trained on the ImageNet-1K dataset~\cite{deng2009imagenet}, and instantiate each architecture in all four numerical formats.
To assess fault tolerance, we subject the resulting models to experimental EMFI campaigns using a NewAE ChipSHOUTER platform and measure the resulting degradation in Top-1 and Top-5 accuracy under identical attack conditions.
Beyond the accuracy evaluation, we also characterize the injected fault pattern itself by analyzing the bit error rate, the spatial distribution of corrupted bytes, and the prevalence of \texttt{0xFE}/\texttt{0xFF} byte values across formats.
This methodology enables a direct comparison of the robustness of different weight representations and helps explain the underlying mechanisms of failure.
Our results reveal marked differences across formats: although the measured fault density is broadly similar across representations, floating-point models exhibit catastrophic degradation due to NaN-inducing exponent corruption, whereas integer models generally retain substantially higher accuracy, with INT8 offering the most favorable trade-off between baseline accuracy and EMFI resilience.
To our knowledge, this is the first comparative study of neural network fault-injection robustness across multiple weight precisions under a unified experimental setup.

\textbf{Organization.} The rest of the paper is organized as follows. 
Section~\ref{sec:related} overviews a related work in the field and Section~\ref{sec:background} gives the necessary background.
Evaluation method on an experimental hardware is described in Section~\ref{sec:method}, followed by Section~\ref{sec:results} that details the results of this work.
Section~\ref{sec:discussion} provides the discussion, and finally, Section~\ref{sec:conclusion} concludes this paper and provides directions for future work.

\begin{table*}[ht]
\centering
\caption{Summary of related work on fault injection on neural network representations.}
\begin{tabular}{|l|p{3cm}|p{4cm}|p{2.5cm}|p{4cm}|}
\hline\small
\textbf{Prior work} & \textbf{Attack type} & \textbf{Target model \& platform} & \textbf{Weight format} & \textbf{Main findings} \\
\hline\hline
Hong \textit{et al.}~\cite{hong2019terminal} & Bit-flip attack (memory fault via Rowhammer) & 19 Deep CNN classifiers; simulated attack & 32-bit floating point & Flipping a specific single bit can cause an accuracy loss of over 90\%.\\
\hline
Rakin \textit{et al.}~\cite{rakin2019bitflip} & Bit-flip attack (memory fault via Rowhammer) & Deep CNN classifiers; simulated attack & 8/6/4-bit integer & Flipping $\sim$13 weight bits can degrade a network’s accuracy to $\sim$0\% (misclassification of virtually all inputs). \\
\hline
Breier \textit{et al.}~\cite{breier2021sniff} & Fault injection for model extraction (sign-bit flips) & Deep feature-extractor network; simulated attack & 32/64-bit floating point & By flipping sign bits of activations, achieved $<10^{-13}$ error in recovered weights, effectively stealing the DNN model. \\
\hline
Libano \textit{et al.}~\cite{libano2020quantization} & Radiation-induced soft errors (fault injection \& accelerated radiation tests) & Small CNN on FPGA (MNIST digit recognition) & 1-bit binary vs. 32-bit float & Binary-weight network showed 39\% fewer erroneous outputs under faults, but 12\% more of its errors were catastrophic (misclassifications) compared to full-precision. \\
\hline
Guillemé \textit{et al.}~\cite{guilleme2025sfi4nn} & Statistical fault injection (simulated bit flips) & AlexNet-like CNN, software simulation; evaluated pruning and quantisation & 8-bit integer (with varying pruning levels) & Pruning increases fault sensitivity (less redundancy). Proposed selective redundancy (hardware voting) improved fault tolerance by $\sim$96\% vs. no protection. \\
\hline
Gaine \textit{et al.}~\cite{gaine2023fault} & Electromagnetic \& laser fault injection (instruction skip) & CNN (2-layer ConvNet) on ARM Cortex-M4 microcontroller (embedded device) & 8-bit integer (CMSIS-NN quantized) & First demonstrated EMFI/LFI on a real CNN inference: single-glitch instruction skips in convolution or activation routines caused targeted mispredictions and persistent erroneous states. \\
\hline
Goswami \textit{et al.}~\cite{goswami2025emfi} & Electromagnetic fault injection (transient bit flips in NVM) & TinyML hardware platform with FRAM storage; models: MobileNet, ResNet, EfficientNet on CIFAR-10 & 8-bit integer & EM pulses during weight loading corrupted stored weights, leading to up to 40\% accuracy loss in lightweight models (larger models fared better due to higher redundancy). \\
\hline
\textbf{This work} & Electromagnetic fault injection (transient bit flips in NVM) & Embedded memory chip hosting model parameters of ResNet-18, ResNet-34, ResNet-50, VGG-11 & 32/16-bit floating point, 8/4-bit integer & Floating point representations degrade significantly in accuracy, while integer ones offer more resistance to faults, especially for larger networks. The best trade-off is offered by the 8-bit integer quantization.\\
\hline
\end{tabular}
\label{tab:related_work}
\end{table*}

\section{Related Work}
\label{sec:related}
A number of prior works have begun to explore the intersection of neural network quantization/representation and fault attacks, which we briefly review here and summarize in Table~\ref{tab:related_work}. 

One of the first attacks showed that by selecting and flipping a specific bit in a 32-bit floating point representation, it is possible to reduce the network accuracy by $\approx$90\%~\cite{hong2019terminal}.

In the domain of fault attacks on quantized neural networks, one of the earliest studies showed that by strategically flipping a few weight bits (via a Rowhammer-induced memory fault), an attacker could force a DNN classifier into misclassifying virtually all inputs, effectively reducing its accuracy to nearly random guess~\cite{rakin2019bitflip}.
Along with the previous work, this demonstrated the outsized impact that even low-level data corruption can have on network accuracy, whether it is represented in a floating-point or an integer format.

SNIFF method~\cite{breier2021sniff} took fault attacks further by using them for model extraction: it allowed to reverse-engineer the parameters of a neural network by inducing sign-bit flips in the network’s computations. 
By using high-precision arithmetic in their experiments, the authors were able to recover model weights with negligible error (on the order of $10^{-13}$), essentially stealing a proprietary model with far fewer queries than a standard black-box approach. 

On the other hand, research in neural network reliability has looked at how model design choices (like quantization) affect tolerance to unintentional faults (for example, soft errors or radiation-induced bit flips). 
In~\cite{libano2020quantization}, the authors investigated CNNs on FPGAs under radiation-induced single-event upsets, comparing a baseline full-precision model to a binarized (1-bit weights) model. 
They found that the quantized (binary) network was about 39\% less sensitive to radiation-induced faults than the full-precision network, presumably because the simpler binary representation and reduced parameter count offered a smaller attack surface in terms of bits that can be flipped. 
However, they also observed a trade-off: the fraction of faults that led to output misclassification was slightly higher (by $~\approx$12\%) in the binary network, indicating that while fewer faults occured, those that did were somewhat more likely to cause an error in inference. 

In another recent study~\cite{guilleme2025sfi4nn}, the authors examined the fault tolerance of quantized and pruned CNN models. 
They developed a statistical fault injection framework (SFI4NN) to efficiently evaluate large networks, and reported that as networks are more aggressively pruned (sparser), their sensitivity to bit flips increases, presumably due to less redundancy in the parameters.
To counter this, they proposed a selective redundancy mechanism (a lightweight error mitigation technique) and demonstrated it could improve a quantized CNN’s resilience by about 96\% compared to an unprotected model.

On the hardware side, \cite{gaine2023fault} recently demonstrated practical EMFI and laser fault attacks on a microcontroller running an 8-bit quantized CNN. 
They showed that even a single instruction skip fault (induced by an EM pulse or laser shot) during a forward pass can alter the network’s prediction or cause a persistent ``memory effect'' in the computations, highlighting the real-world feasibility of such attacks on TinyML devices. 

Another closely related work~\cite{goswami2025emfi} specifically looked at EMFI on a TinyML hardware platform with INT8 weight storage. 
They found that small, lightweight models suffered major accuracy degradation under EMFI, whereas larger models with more parameters (and presumably more redundancy) were somewhat more robust. 

In summary, these related studies show a growing awareness of the security implications of neural network representations. 
None of the existing works, however, provides a direct head-to-head comparison of different numeric formats under the same fault injection conditions and with the same models -- the gap that our paper aims to fill. 
By comparing FP32, FP16, INT8, and INT4 in a unified experimental setup, our work builds upon the above research to provide new insights into how model precision and fault resilience are connected.

\section{Background}
\label{sec:background}
In this section, we first provide an overview of embedded neural network implementations in Subsection~\ref{sec:embeddedNN}.
We then describe the number formats used for weight representations in this work in Subsection~\ref{sec:numbers}.
Finally, we discuss the attack goals of FIAs on neural network implementations in Subsection~\ref{sec:FIAonNN} and the physical principles underlying electromagnetic fault injection in Subsection~\ref{sec:emfi}.

\subsection{Embedded and Edge Neural Network Implementations}
\label{sec:embeddedNN}
The deployment of deep learning models on edge devices, often referred to as Edge AI or TinyML~\cite{alajlan2022tinyml}, represents a paradigm shift from centralized cloud-based computation to on-device inference.
This shift is driven by the need for low latency, improved data privacy, and reduced bandwidth consumption.
However, embedded targets, ranging from constrained microcontrollers (MCUs) to embedded GPUs, operate under strict power and memory budgets.
This necessitates the use of specialized software frameworks designed to support the numerical representations described in Subsection~\ref{sec:numbers}.

\subsubsection{TinyML and Resource Constraints}
TinyML specifically targets the most resource-constrained class of devices, typically based on microcontrollers (e.g., ARM Cortex-M) with limited SRAM and flash memory, and often operating in the milliwatt power range.
Unlike server-grade deployments, such implementations cannot rely on massive parallelism or virtually unlimited memory.
Consequently, models must be aggressively compressed and optimized before deployment.
In this context, the physical integrity of memory is paramount; investigating fault injection in TinyML is therefore critical, because the redundancy typically present in large, over-parameterized cloud models is often removed during the optimization process.

\subsubsection{TensorFlow Lite (TFLite)}
TensorFlow Lite~\cite{david2021tensorflow} is a widely adopted open-source framework for on-device inference.
Its deployment workflow involves converting a trained model into the efficient FlatBuffer-based \texttt{.tflite} format using the TFLite Converter.
This conversion process performs several graph optimizations that are essential for embedded execution:
\begin{itemize}
    \item \textbf{Operator fusion:} combining multiple operations (e.g., convolution followed by an activation function) into a single computational kernel to reduce memory-access overhead.
    \item \textbf{Quantization-aware conversion:} transforming weights from FP32 to INT8 or INT4, thereby allowing the model to run on integer-only arithmetic logic units (ALUs) commonly found in low-power MCUs.
\end{itemize}
The TFLite Interpreter then executes the optimized graph.
Because TFLite relies on static memory planning to minimize runtime allocation, it presents a deterministic target for electromagnetic fault-injection attacks.

\subsection{Number Formats}
\label{sec:numbers}
In this study, we consider two floating-point representations and two integer representations (see Table~\ref{tab:representations} for a summary).
In the following, we briefly describe each representation; additional background can be found in~\cite{parhami1999computer}.

\subsubsection{Floating-Point Representations}
Floating-point arithmetic provides a wide dynamic range and is therefore the standard representation for training deep learning models.
These formats follow the IEEE~754 standard, dividing the binary representation into three components: the sign bit ($s$), the exponent bits, and the fraction or mantissa ($m$). The exponent bits are the binary representation of the biased exponent $e$.

\paragraph{Single Precision (FP32)}
FP32 is the standard format for training neural networks on general-purpose GPUs and CPUs.
It uses \(32\) bits allocated as follows: \(1\) sign bit, 8 exponent bits, and \(23\) mantissa bits.
The value $v$ of an FP32 number is given by Equation~(\ref{eq:fp32}):

\begin{equation}
    v_{FP32} = (-1)^s \times 2^{e - 127} \times (1.m)
    \label{eq:fp32}
\end{equation}

This format provides a high dynamic range (approximately $\pm 3.4 \times 10^{38}$) and high precision, making it robust against vanishing gradients during training, but memory-intensive for inference on embedded targets.

\paragraph{Half Precision (FP16)}
To reduce memory bandwidth and storage requirements, FP16 compresses the representation into \(16\) bits: \(1\) sign bit, \(5\) exponent bits, and \(10\) mantissa bits.
Its value is given by Equation~(\ref{eq:fp16}):

\begin{equation}
    v_{FP16} = (-1)^s \times 2^{e - 15} \times (1.m)
    \label{eq:fp16}
\end{equation}

While FP16 halves the memory footprint compared with FP32, the smaller number of exponent bits significantly shrinks the dynamic range (approximately $\pm 6.5 \times 10^4$).
This makes the representation more sensitive to numerical instability, although it is often sufficient for inference in many embedded applications.

\begin{figure*}[tb]
    \centering
    \includegraphics[width=0.8\linewidth]{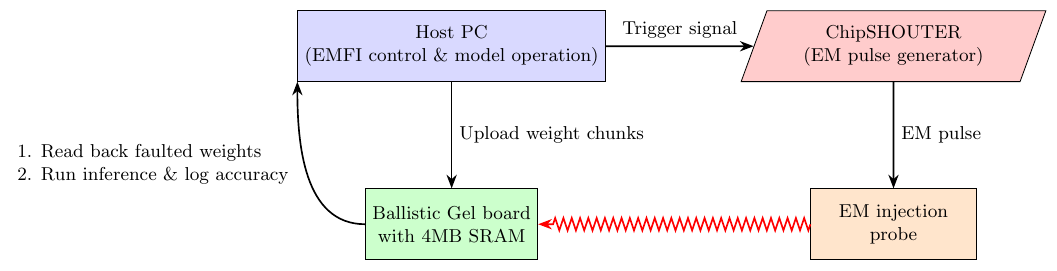}
    \caption{Experimental electromagnetic fault injection setup overview.}
    \label{fig:setup_overview}
\end{figure*}

\subsubsection{Quantized Integer Representations}
Quantization maps continuous floating-point values to a discrete set of low-precision integer values.
This is critical for embedded inference, because integer arithmetic units are generally faster and more energy-efficient than floating-point units.
In uniform affine quantization, a real-valued quantity $r$ is mapped to a quantized integer $q$ using a scale factor $S$ and a zero-point $Z$:
\begin{equation}
    q = \operatorname{round}\!\left(\frac{r}{S}\right) + Z
    \label{eq:quant}
\end{equation}

\paragraph{8-bit Integer (INT8)}
INT8 is the prevalent quantization standard for edge AI inference (e.g., TensorFlow Lite and TensorRT).
It uses \(8\) bits to represent signed integers, typically in two's complement notation, resulting in a range of $[-128, 127]$.
INT8 reduces the model size by a factor of \(4\) compared with FP32.
Because its dynamic range is limited to \(256\) discrete levels, INT8 typically relies on uniform affine quantization, in which floating-point weights are mapped to integers using calibrated quantization parameters \(S\) and \(Z\) to minimize accuracy degradation.

\paragraph{4-bit Integer (INT4)}
INT4 represents an aggressive low-bit-width format used in extreme edge scenarios or for compressing large language models (LLMs).
It uses only \(4\) bits per weight, providing an \(8\times\) compression ratio compared with FP32, with a signed two's complement range of $[-8, 7]$.
However, having only \(16\) representable values makes this format highly sensitive to perturbations.
In the context of fault injection, a single bit flip in an INT4-encoded weight can induce a large relative change in its numerical value, corresponding to a substantial fraction of the total dynamic range.

\begin{table}[htbp]
\caption{Summary of the numerical representations considered in this work.}
\begin{center}
\begin{tabular}{|c|c|c|c|}
\hline
\textbf{Format} & \textbf{Total bits} & \textbf{Range} & \textbf{Decimal digits} \\
\hline
FP32 & 32 & $\pm 3.4 \times 10^{38}$ & $\sim$7 \\
FP16 & 16 & $\pm 6.55 \times 10^{4}$ & $\sim$3--4 \\
INT8 & 8 & $[-128, 127]$ & Exact \\
INT4 & 4 & $[-8, 7]$ & Exact \\
\hline
\end{tabular}
\label{tab:representations}
\end{center}
\end{table}

\subsection{Fault Injection Attacks on Neural Networks}
\label{sec:FIAonNN}
Fault injection attacks (FIA) were originally proposed as an active hardware attack method on cryptosystems, allowing easy key extraction even in cases where cryptanalysis attack would have been impractical~\cite{boneh2001importance,breier2019automated}.
FIAs on Deep Neural Networks (DNNs) represent a critical security threat where an adversary manipulates the model's output by inducing transient or permanent errors in the underlying hardware \cite{liu2017fault}. 
In the context of edge AI, these attacks typically target the model weights stored in memory~\cite{rakin2019bitflip} or the activations during runtime computation~\cite{hou2021physical,breier2018practical}.
Such corruptions allow several different attack vectors:
\begin{itemize}
    \item Misclassification: the basic attack, referred to as ``evasion'' in the context of adversarial learning. It can either be untargeted~\cite{rakin2019bitflip} or targeted~\cite{rakin2021t}, meaning that the attacker either achieves a random class output or a specific class output.
    \item Reverse engineering: also referred to as ``model stealing/extraction.'' It was shown that precisely controlled bit flips can precisely recover model parameters~\cite{breier2021sniff}.
    \item Backdoor/trojan planting: in this attack, the attacker either plants the backdoor during the training~\cite{breier2022foobar,martinez2024deepbar} or after the model deployment~\cite{rakin2020tbt}. Then, a specific trigger inserted in the network input causes the model misbehavior.
\end{itemize}
Additionally, fault attacks can cause a generic denial of service attack by making the system unresponsive, but this attack vector is not specific to neural networks and can generally be achieved by simpler means in practice (e.g., by making the power source unavailable).

\subsection{Electromagnetic Fault Injection}
\label{sec:emfi}
Electromagnetic fault injection (EMFI) is a non-invasive physical attack technique used to disrupt the normal execution of an integrated circuit (IC) by inducing localized transient faults~\cite{habibi2021emfi}. 
Unlike contact-based methods such as clock or voltage glitching, EMFI utilizes Faraday's Law of Induction to create internal currents without physical modification of the target package.
While it can be more precise if the IC package is opened, faults can still be triggered through the package, making it one of the biggest advantages over the laser fault injection~\cite{breier2015testing}, along with the equipment cost~\cite{breier2022practical}.

The core principle of EMFI involves a high-voltage pulse generator connected to an injection probe, typically consisting of a copper coil wound around a ferrite or mu-metal core~\cite{hou2024cryptography}. 
When a rapid current pulse passes through the coil, it generates a time-varying magnetic field $\vec{B}(t)$. 
According to Faraday’s Law (see, for example,~\cite{kuehn2015faraday}), the electromotive force (EMF) induced in the underlying silicon circuitry is given by:

\begin{equation}
    \mathcal{E} = -\frac{d\Phi_B}{dt} = -\frac{d}{dt} \int \vec{B} \cdot d\vec{A}
\end{equation}

where $\Phi_B$ is the magnetic flux passing through the sensitive loops of the IC's metal layers. 
This induced voltage can momentarily exceed the threshold voltages of transistors or disrupt the propagation of signals along data buses, leading to bit-flips in registers, memory cells, or instruction flow errors~\cite{dumont2019electromagnetic}.

The efficacy of an EMFI attack is primarily governed by the rise time, pulse width, and peak voltage of the injection signal. 
In our setup, we utilize a high-voltage pulse generator (NewAE ChipSHOUTER) capable of generating transient pulses with a magnitude of 200-500V.

\textbf{Injection probes.} The injection probe acts as the transducer that converts the electrical pulse into a localized magnetic field. 
Our study utilizes a micro-probe designed for high-precision spatial targeting.

The following properties influence the success and the precision of the EMFI attack~\cite{gaine2022new}:
\begin{itemize}
    \item Core material: the probe utilizes a high-permeability ferrite core. The ferrite material serves to concentrate the magnetic flux lines, minimizing far-field interference and ensuring that the fault is localized to a specific region of the SoC (e.g., the SRAM block or the instruction pipeline).
    \item Coil topology: the probe consists of several turns (typically 5--20) of fine copper wire. A lower number of turns reduces the total inductance, allowing for faster rise times, whereas a higher number of turns increases the peak magnetic field strength at the cost of temporal resolution.
    \item Probe tip geometry: the probe tip determines the shape and the size of the electromagnetic field generated by the probe. We use a relatively large tip with a diameter of 4 mm, allowing us to target multiple memory banks at the same time.
\end{itemize}
To characterize the vulnerability of the target, the probe is normally mounted on a high-precision XYZ motorized stage. 
By automating the displacement of the probe across the surface of the IC, a ``fault map'' can be generated for further analysis.
In our case, the mapping process was synchronized with the trigger signal of the DNN inference, ensuring that the pulse was injected during the exact window when the weights of the targeted layers were stored in the memory.

\section{Evaluation Method}
\label{sec:method}
In this section, we first present the experimental setup in Subsection~\ref{sec:setup}, followed by a discussion of the surface scan results, which identify the most vulnerable region of the chip, in Subsection~\ref{sec:scan}.
The quantization configuration is detailed in Subsection~\ref{sec:configuration}.

\begin{figure}[htb]
    \centering
    \includegraphics[width=0.6\linewidth]{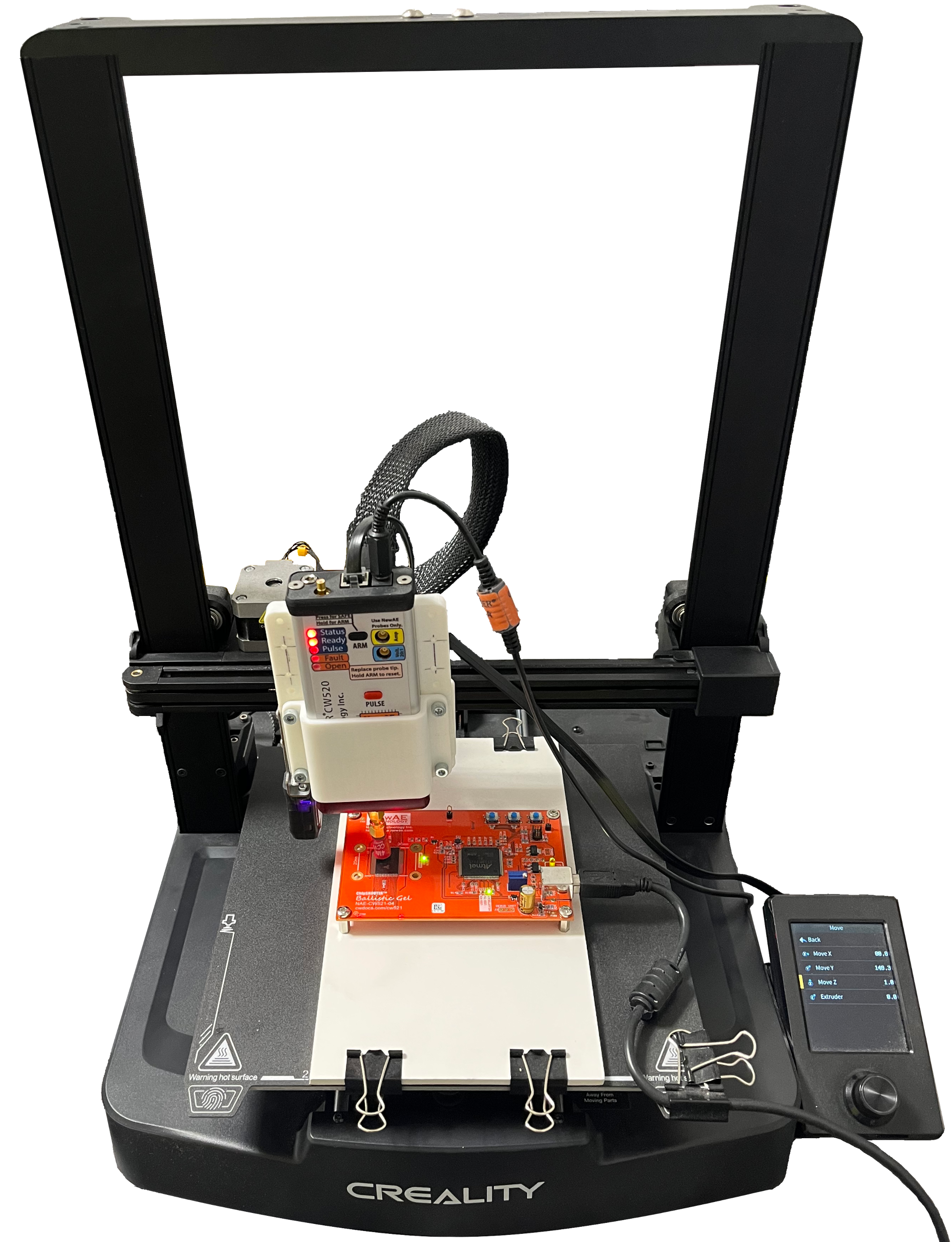}
    \caption{NewAE ChipSHOUTER EMFI device mounted on the Ender-3 3D printer used as a positioning device. The Ballistic Gel board with the 4MB SRAM chip used as the DUT is positioned below the ChipSHOUTER.}
    \label{fig:chipshouter}
\end{figure}

\begin{figure}[htb]
    \centering
    \includegraphics[width=0.6\linewidth]{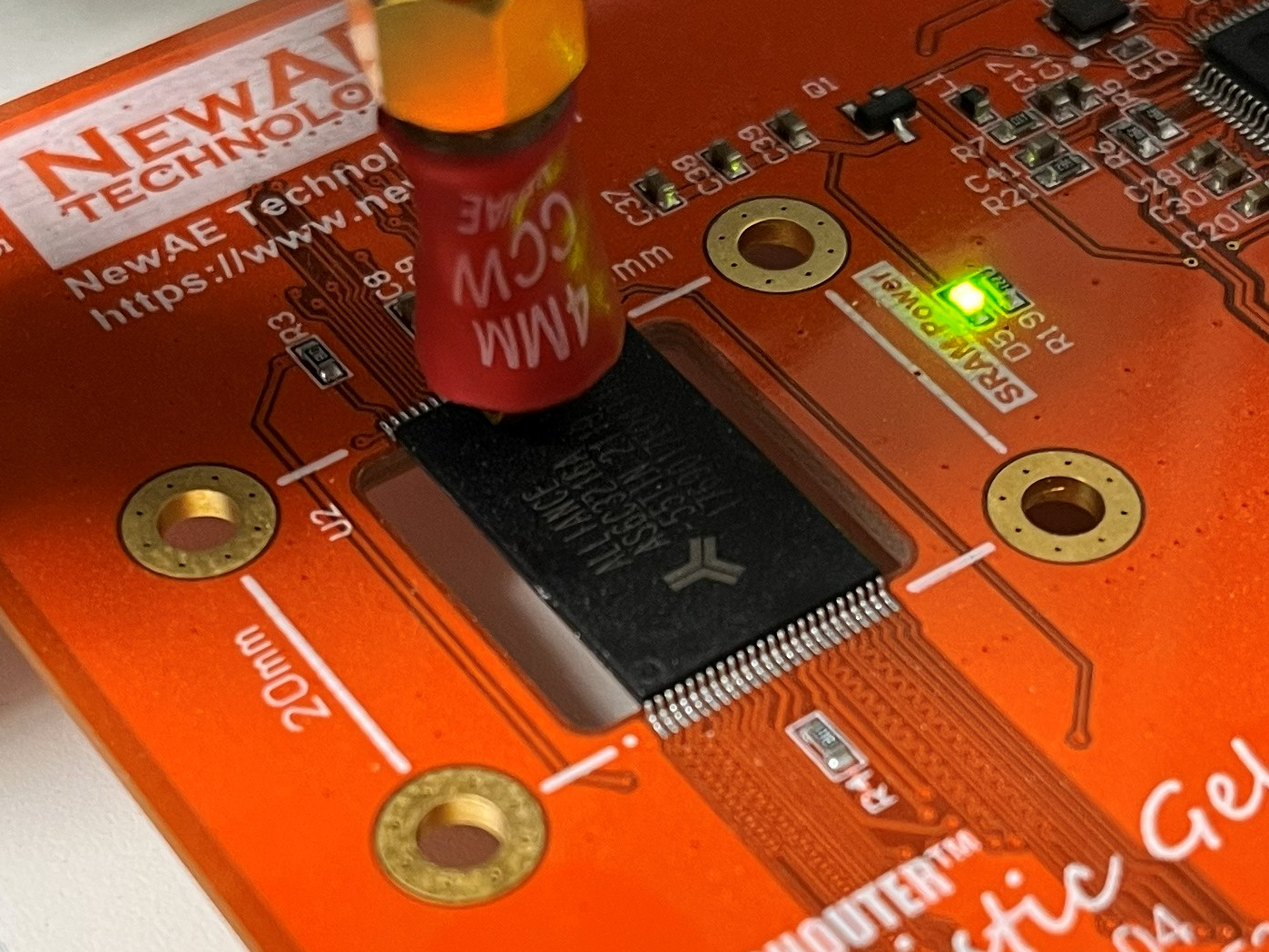}
    \caption{Detail of the EM probe above the SRAM chip.}
    \label{fig:probe}
\end{figure}

\subsection{Experimental Setup}
\label{sec:setup}
The components of the experimental setup and their respective functions are shown in Fig.~\ref{fig:setup_overview}.
The setup is centered around the NewAE ChipSHOUTER electromagnetic fault injection device\footnote{\url{https://chipwhisperer.readthedocs.io/en/latest/ChipSHOUTER/ChipSHOUTER.html}}.
A key component of the EMFI setup is the injection probe.
We used a \(4\)\,mm probe with counter-clockwise winding around its ferrite core (see Fig.~\ref{fig:probe} for details).
Using this setup, we targeted the CW521 Ballistic Gel board, which hosts a \(4\)\,MB SRAM chip (AS6C3216A-55TIN, ALLIANCE MEMORY) fabricated in CMOS technology.
In the remainder of the paper, we refer to this board as the DUT (device under test).
An Ender-3 V3 SE 3D printer was repurposed as an XYZ positioning platform by replacing the printhead with a ChipSHOUTER mounted on a custom 3D-printed holder.
The resulting fault-injection bench is shown in Fig.~\ref{fig:chipshouter}.
The Host PC controlled the ChipSHOUTER and the positioning platform, communicated with the DUT, and handled all operations related to the neural network model.

After the initial evaluation, we identified a set of fault injection parameters that reliably produced memory faults.
Specifically, we used a 300\,V pulse with a width of \(160\)\,ns.
During each attack, a single electromagnetic pulse was injected.

This experimental architecture reflects an in-storage fault-injection threat model, in which the adversary has physical access to the weight-storage memory before or during device boot (e.g., via proximity to external SRAM on a PCB-level device), which is an established scenario in embedded systems security~\cite{gaine2023fault}.
Performing inference on a separate, unmodified GPU allows us to isolate the mathematical propagation of representation-level faults from confounding effects in the inference stack, such as memory interleaving, instruction-level caching, and ECC mechanisms commonly present in integrated SoCs.
Extending this analysis to fully in-situ inference on a microcontroller is an important direction for future work, but it lies outside the scope of the present study, whose objective is the comparative characterization of sensitivity across numerical representations.

\subsection{Surface Scan}
\label{sec:scan}
\begin{figure}[htb]
    \centering
    \includegraphics[width=0.8\linewidth]{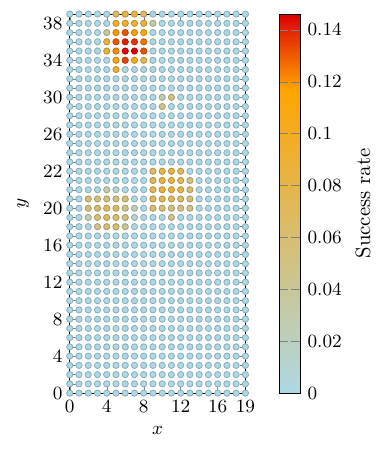}
    \caption{Surface scan showing the success rate at each point, expressed as the proportion of corrupted bytes in the SRAM.}
    \label{fig:scan}
\end{figure}

To identify the probe position that most effectively induced faults, we first performed a surface scan of the chip.
The AS6C3216A-55TIN is packaged in a TSOP I surface-mount package with dimensions of $12 \times 20$~mm$^2$.
The Z-axis was fixed at a height of \(1\)~mm above the chip surface.
We used \(20\) sampling points along the \(X\)-axis and \(40\) along the \(Y\)-axis, corresponding to a step size of approximately \(0.5\)~mm and yielding a total of 800 scan points.
At each point, we uploaded random bytes to fill the entire SRAM and then performed electromagnetic fault injection.
After the injection, we read back the memory contents and compared them with the original data to determine the proportion of corrupted bytes, which we define as the success rate.
The resulting scan is shown in Fig.~\ref{fig:scan}.
The most sensitive region is located around coordinate $(7,35)$, where the success rate reaches approximately \(15\%\).
The position of the probe above this region is shown in Fig.~\ref{fig:probe}.

\subsection{Quantization Configuration}
\label{sec:configuration}
All integer models were quantized using Brevitas PTQ (v0.11, \texttt{Xilinx/brevitas})~\cite{brevitas}.
Post-training quantization was performed using symmetric weight quantization (\texttt{weight-quant-type: sym}), per-channel granularity (\texttt{weight-quant-granularity: per\_channel}), and integer format (\texttt{quant-format: int}).
To preserve accuracy under aggressive \(4\)-bit quantization, GPTQ-style optimization was enabled (\texttt{gptq: true}, \texttt{gpxq\_act\_order: true}), which explains the relatively small clean-accuracy degradation of INT4 on ImageNet (cf.\ Table~\ref{tab:vanilla}).

\textbf{INT4 packing.}
Nibble pairs are stored in ascending address order.
Integer values $v_1, v_2, v_3, v_4, \ldots$ are packed into bytes as $\texttt{0x}v_1v_2,\,\texttt{0x}v_3v_4,\ldots$ using little-endian nibble order and signed two's complement representation over the range $[-8,\,7]$.
Consequently, a single byte-corruption event simultaneously invalidates two adjacent \(4\)-bit weights.

\section{Experimental Results}
\label{sec:results}
To evaluate the impact of EMFI on model integrity, we used the ImageNet-1K validation set, sub-sampled to \(4,096\) images with a fixed random seed (\(21\)) to ensure reproducibility.
Because ImageNet-1K comprises \(1,000\) classes, random guessing corresponds to an accuracy of \(0.1\%\).
Inference was performed using ONNX Runtime with CUDA acceleration on an NVIDIA RTX 4080 SUPER.
Model degradation was evaluated using Top-1 and Top-5 accuracy across four weight representations: FP32, FP16, INT8, and INT4.
The vanilla accuracies of all evaluated models are reported in Table~\ref{tab:vanilla}.

\begin{table}[!t]
\centering
\caption{Vanilla Top-1 and Top-5 accuracy (\%) of the evaluated architectures across four weight representations.}
\label{tab:vanilla}
\small
\begin{tabular}{|llcc|}
\hline
\textbf{Model} & \textbf{Type} & \textbf{Top-1 (\%)} & \textbf{Top-5 (\%)} \\ \hline
\multirow{4}{*}{ResNet18} & FP32 & 68.750 & 88.379 \\
                          & FP16 & 68.774 & 88.379 \\
                          & INT8 & 68.774 & 88.452 \\
                          & INT4 & 67.236 & 87.549 \\ \hline
\multirow{4}{*}{ResNet34} & FP32 & 72.778 & 90.991 \\
                          & FP16 & 72.803 & 90.967 \\
                          & INT8 & 72.583 & 90.942 \\
                          & INT4 & 71.973 & 90.308 \\ \hline
\multirow{4}{*}{ResNet50} & FP32 & 75.562 & 92.456 \\
                          & FP16 & 75.513 & 92.481 \\
                          & INT8 & 75.244 & 92.334 \\
                          & INT4 & 74.683 & 91.821 \\ \hline
\multirow{4}{*}{VGG11}    & FP32 & 68.311 & 88.623 \\
                          & FP16 & 68.335 & 88.599 \\
                          & INT8 & 68.189 & 88.648 \\
                          & INT4 & 68.164 & 88.062 \\ \hline
\end{tabular}
\end{table}

We first characterize the injected fault pattern in Subsection~\ref{sec:characterization}, then quantify the floating-point-specific corruption effects in Subsection~\ref{sec:fp_corruption}. 
Finally, Subsection~\ref{sec:main_result} reports the post-EMFI accuracy across all architectures and weight representations.

\subsection{Characterization of the EMFI Error Pattern}
\label{sec:characterization}
To quantify the impact of EMFI on the target memory subsystem, we first characterized the fault effects on a linear data buffer.
By applying a bitwise XOR operation between the original and post-attack data, we mapped the distribution of corrupted bits across the logical address space.

\begin{figure}[htb]
    \centering
    \includegraphics[width=0.8\linewidth]{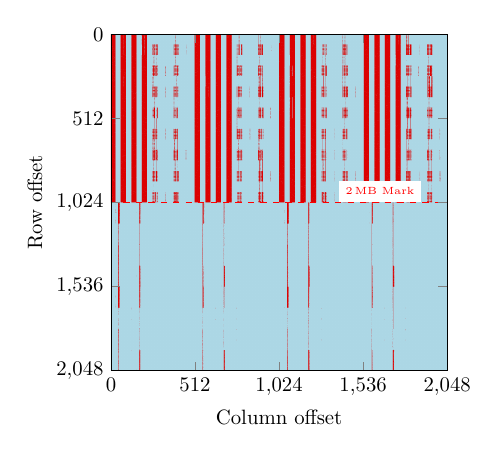}
    \caption{Logical memory map of EMFI effects on a \(4\)\,MB weight buffer. Red indicates corrupted memory locations. A significant decrease in fault density is observed after the \(2\)\,MB offset, suggesting either a boundary transition in the memory controller's handling of burst data or a physical proximity limit of the EM probe.}
    \label{fig:logical_map}
\end{figure}

As shown in Fig.~\ref{fig:logical_map}, the faults exhibit a distinct periodic pattern.
Although the map is visualized as a two-dimensional matrix to highlight recurring structures, it represents a linear sequence of addresses rather than the physical topology of the RAM cells.
A significant transition is observed at the \(2\)\,MB offset:
\begin{itemize}
    \item \textbf{\(0\)--\(2\)\,MB region:} a high density of periodic corruption, suggesting a specific interaction between the EM pulse and the memory controller's burst-access logic or internal buffering.
    \item \textbf{Post-\(2\)\,MB region:} a sharp decline in fault density, which may indicate either a boundary in the memory allocation or a physical distance limit of the EM probe's effective field relative to the underlying chip.
\end{itemize}
This pattern proved highly deterministic and persisted across power cycles and multiple days of testing.
Such repeatability suggests that the vulnerability is not a stochastic ``white-noise'' effect, but is instead tied to the architectural handling of data at specific logical offsets.

\begin{table}[t]
\centering
\caption{Bit error rate per weight format (mean\,$\pm$\,std across all
  fault-injected chunks and architectures).}
\label{tab:ber}
\begin{tabular}{|l|c|c|}\hline
\textbf{Format} & \textbf{Mean BER (\%)} & \textbf{Std BER (\%)} \\
\hline
FP32  & 5.97 & 0.36 \\
FP16  & 6.07 & 0.72 \\
INT8  & 7.51 & 0.51 \\
INT4  & 6.15 & 1.30 \\
\hline
\end{tabular}
\end{table}

\begin{figure}[htb]
    \centering
    \includegraphics[width=0.98\linewidth]{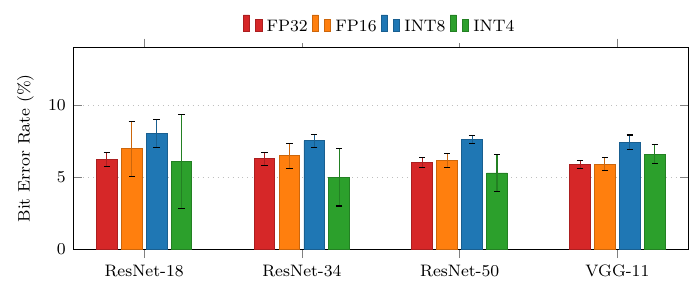}
    \caption{Bit error rate (BER) per architecture and weight representation.
           Error bars indicate $\pm 1$ standard deviation across all
           fault-injected chunks. BER is largely format-independent
           (${\approx}6\text{--}8\%$ across all configurations), confirming
           that the observed accuracy differences stem from the
           representation's response to faults rather than from differences
           in injected fault density.}
    \label{fig:ber}
\end{figure}

\textbf{Bit error rate.}
To quantify the fault density across all \(352\) fault-injected chunks, we computed the bit error rate (BER) as the ratio of changed bits to the total number of bits in the \(4\)\,MB window.
Table~\ref{tab:ber} and Fig.~\ref{fig:ber} summarize the results.
Across all models and formats, the mean BER lies within the narrow range of $5.97$--$7.51\%$, with particularly low standard deviations for FP32 ($\pm 0.36\%$) and INT8 ($\pm 0.51\%$), confirming a stable and reproducible fault pattern.
Crucially, the BER is largely format-independent, with all four representations exhibiting statistically similar fault densities.
This indicates that the pronounced accuracy differences reported in this section arise from the representations' responses to the injected faults, rather than from differences in the number of faults induced.

\begin{figure}[htb]
    \centering
    \includegraphics[width=0.98\linewidth]{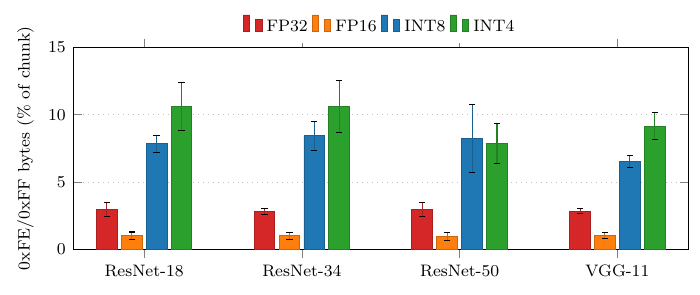}
    \caption{Proportion of bytes taking the value \texttt{0xFE} or \texttt{0xFF} per injected chunk.
           Although integer formats receive a \emph{higher} fraction of
           \texttt{0xFE}/\texttt{0xFF} bytes than floating-point formats, they exhibit far
           lower accuracy degradation.}
    \label{fig:feff}
\end{figure}

\textbf{0xFE/0xFF byte pattern.}
Fig.~\ref{fig:feff} shows the proportion of corrupted bytes that take the value \texttt{0xFE} or \texttt{0xFF}.
For FP32 and FP16, this fraction is modest (${\approx}2.9\%$ and $1.0\%$ of the chunk, respectively), whereas for INT8 and INT4 it is considerably higher ($7.1\%$ and $9.2\%$, respectively).
Despite receiving a \emph{higher} proportion of maximal-value bytes, the integer models exhibit substantially smaller accuracy losses.
This asymmetry is explained by the IEEE\,754 encoding: in FP32, a faulty byte value of \texttt{0xFF} in the biased exponent can drive the number to NaN (Not a Number) or $\pm\infty$ regardless of the mantissa, so even a relatively small absolute number of such faulty bytes is sufficient to poison floating-point inference.
In bounded integer formats, by contrast, \texttt{0xFF} simply clamps the weight to its maximum representable value, which is bounded by the quantization scale and therefore cannot trigger the catastrophic cascades described in the later parts.

\begin{figure*}[htb]
    \centering
    \includegraphics[width=1.0\linewidth]{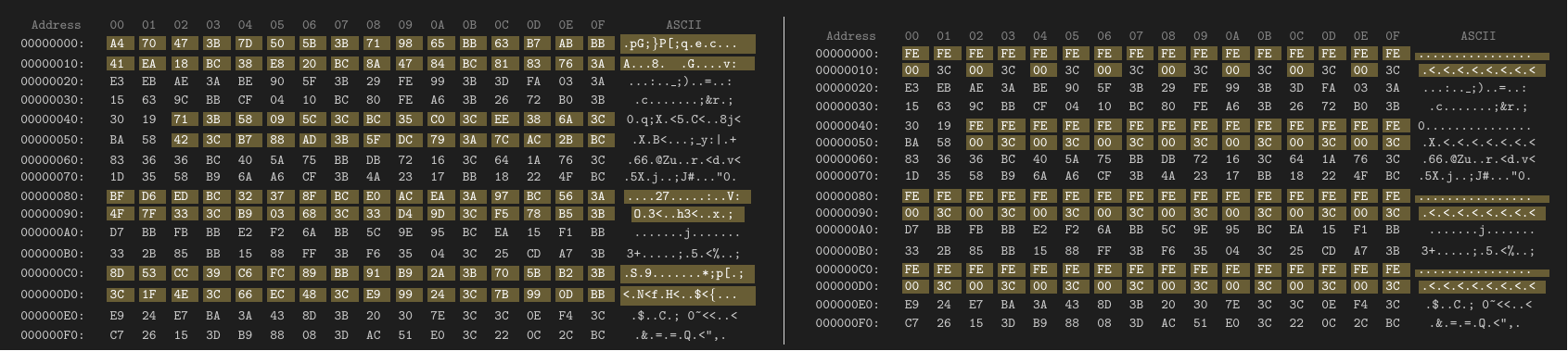}
    \caption{Binary comparison of a weight chunk before (left) and after
    (right) a single EMFI pulse. Entire rows are overwritten with
    \texttt{0xFE}, and alternating rows take the value
    \texttt{0x00\,3C}, consistent with the periodic fault pattern
    observed in Fig.~\ref{fig:logical_map}. In FP32, a byte of \texttt{0xFF}
    or \texttt{0xFE} in the biased exponent field drives the weight to
    NaN or an extremal value regardless of the mantissa.}
    \label{fig:binary}
\end{figure*}

\textbf{Determinism.}
The fault pattern proved highly deterministic: repeated injections at the same \(XY\) position produced the same distribution of corrupted addresses across power cycles and multiple days of testing, consistent with the \(2\)\,MB boundary periodicity observed in Fig.~\ref{fig:logical_map}.
This repeatability confirms that the vulnerability is tied to a systematic interaction between the EM pulse and the memory controller's burst-access logic, rather than to a stochastic noise-floor effect.

\subsection{Floating-Point Corruption Statistics}
\label{sec:fp_corruption}
\begin{figure}[htb]
  \centering
  \begin{minipage}[t]{0.45\linewidth}
    \centering
    \includegraphics[width=\linewidth]{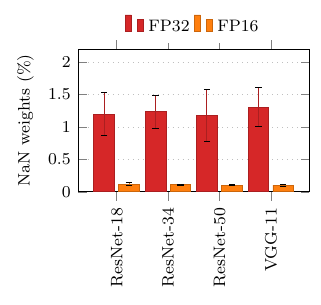}\\[2pt]
    {\footnotesize (a) NaN weight fraction post-attack}
  \end{minipage}%
  \hfill%
  \begin{minipage}[t]{0.53\linewidth}
    \centering
    \includegraphics[width=\linewidth]{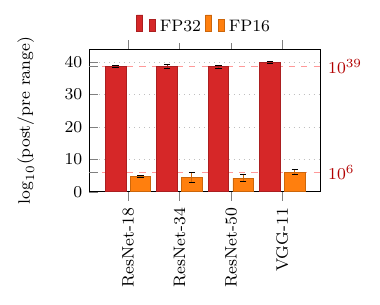}\\[2pt]
    {\footnotesize (b) Weight range expansion ($\log_{10}$ scale)}
  \end{minipage}
  \caption{Floating-point weight corruption statistics per chunk.
           (a)~In FP32, approximately $1.3\%$ of weights per chunk become NaN,
           whereas in FP16 the corresponding fraction is approximately $0.1\%$.
           (b)~The post-attack weight range expands by a factor of
           ${\sim}10^{39}$ for FP32 and ${\sim}10^{6}$ for FP16,
           causing catastrophic activation saturation.
           No $\pm\infty$ values were observed in either format.}
  \label{fig:fp_corruption}
\end{figure}

To quantify the extent of the NaN and range-explosion effects, we extracted three statistics for each floating-point chunk: the fraction of weights that became NaN, and the pre- and post-attack weight ranges.
Table~\ref{tab:fp_corruption} and Fig.~\ref{fig:fp_corruption} summarize the results.

For FP32, between $1.17\%$ and $1.30\%$ of weights per chunk became NaN after a single injection (mean $1.27 \pm 0.32\%$).
No $\pm\infty$ values were observed: the \texttt{0xFF} exponent pattern produced by this EMFI setup generates quiet NaNs rather than infinities.
For FP16, the NaN fraction is approximately $10\times$ lower (${\approx}0.10\%$), owing to the smaller number of exponent bits (\(5\) vs.\ \(8\) bits), which reduces the probability that a random byte-set event produces the all-ones exponent pattern required to encode NaN.

The weight range expands catastrophically in both formats.
The mean range-expansion factor for FP32 is ${\approx}10^{39}$, meaning that the post-attack weight range spans essentially the full representable extent of the format, while FP16 expands by ${\approx}10^{6}$.
A single such extreme weight can propagate through the network's activation functions and saturate downstream neurons, a failure mode that is structurally impossible in INT8 or INT4, where the maximum post-fault weight magnitude is bounded by the quantization scale.

\begin{figure}[tb]
    \centering
    \includegraphics[width=0.98\linewidth]{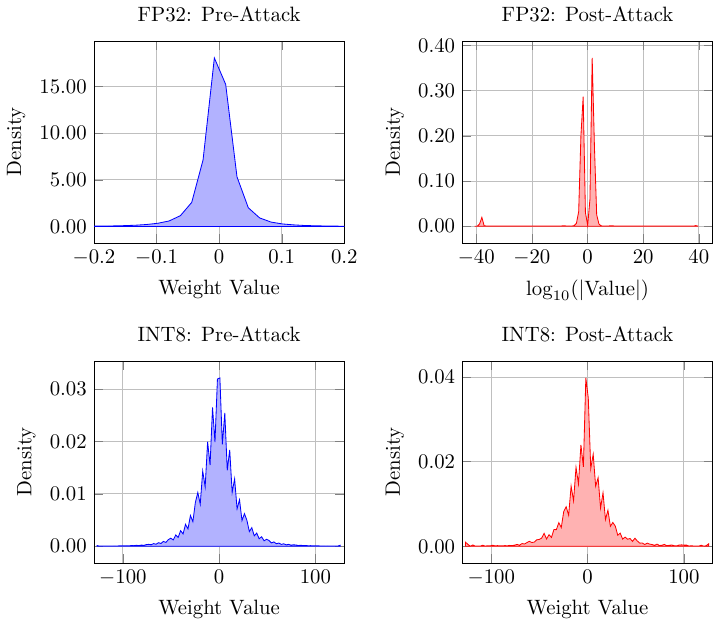}
    \caption{Weight distribution analysis. The FP32 post-attack histogram is shown on a logarithmic \(x\)-axis, $\log_{10}(|\mathrm{Value}|)$, to visualize the ``range explosion'' to $\pm 10^{38}$ caused by EMFI. In contrast, the INT8 pre-attack and post-attack distributions remain within the representable range $[-128,127]$.}
    \label{fig:hist}
\end{figure}

To further investigate the catastrophic failure of floating-point representations relative to the greater resilience of integer formats, we analyzed the weight distributions before and after EMFI.
Fig.~\ref{fig:hist} illustrates the corresponding pre-attack and post-attack histograms for FP32 and INT8.

\textbf{Floating-point explosion.}
As shown in the upper part of Fig.~\ref{fig:hist}, the FP32 weights originally follow a narrow Gaussian-like distribution centered around zero, typically within the range $[-1, 1]$.
However, after a single EM pulse, the weight range expands dramatically.
In FP32, corrupted weights reach extreme magnitudes exceeding $\pm 10^{30}$.
For example, we observed the weight range of a ResNet18 chunk shift from $[-1.04, 0.76]$ to approximately $[-3.4 \times 10^{38}, 3.3 \times 10^{38}]$ post-attack.

This phenomenon is primarily caused by bit flips in the exponent bits.
A flip in the biased exponent changes the scale of the value significantly, producing magnitudes that far exceed those encountered in a normally trained network.
When such extreme values are propagated into subsequent dot-product operations, they quickly induce numerical saturation and the propagation of NaNs throughout the network.
This effectively ``poisons'' the entire inference path, resulting in the near-random accuracy observed in our experiments.

\textbf{Integer bounding.}
In contrast, the lower part of Fig.~\ref{fig:hist} demonstrates the inherent protection provided by quantization.
Although the EM pulse introduces significant noise, visible as the ``flattening'' of the INT8 distribution, the weights remain strictly bounded by the bit-width of the representation (e.g., $[-128, 127]$ for signed INT8).

Even a ``worst-case'' bit flip in an integer format results only in a value clamped to the maximum representable integer of that format.
This prevents the exponential magnitude shifts observed in floating-point formats and ensures that a single corrupted weight cannot dominate the activation of an entire neuron or saturate subsequent layers.
This mathematical bounding is the primary reason why larger models such as VGG-11, which possess a higher degree of parameter redundancy, can maintain high Top-1 accuracy despite high byte-level corruption in memory.

\begin{table}[htb]
\centering
\caption{Floating-point weight corruption statistics
  (mean\,$\pm$\,std across all FP chunks). No Inf values were observed.
  Range expansion = post-attack range / pre-attack range.}
\label{tab:fp_corruption}
\setlength{\tabcolsep}{4pt}
\begin{tabular}{|l|c|c|c|}
\hline
Format & NaN (\%) & Inf (\%) & Range expansion \\
\hline
FP32 & $1.27 \pm 0.32$ & $0$ & $\sim\!10^{39}$ \\
FP16 & $0.10 \pm 0.01$ & $0$ & $\sim\!10^{6}$  \\
\hline
\end{tabular}
\end{table}

\subsection{Sensitivity Analysis across Architectures}
\label{sec:main_result}
\begin{figure*}[tb!]
    \centering
    \includegraphics[width=1.0\linewidth]{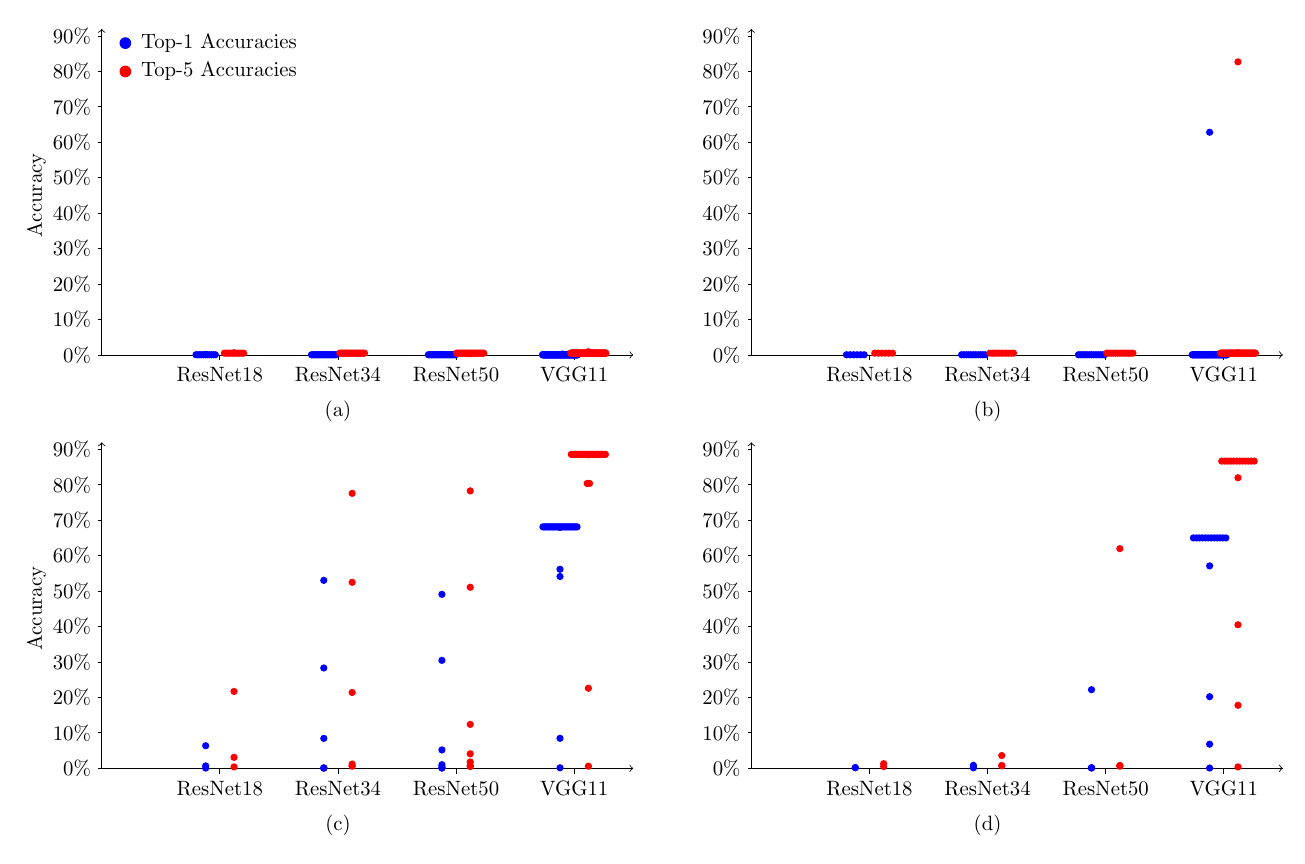}
    \caption{Classification accuracy under fault injection for ResNet-18, ResNet-34, ResNet-50, and VGG-11. Each marker corresponds to the Top-1 (blue) or Top-5 (red) accuracy obtained after injecting faults into a specific chunk of the model's weight parameters. Results are reported for four weight representations: (a) FP32, (b) FP16, (c) INT8, and (d) INT4.}
    \label{fig:accuracies}
\end{figure*}

We conducted a granular sensitivity analysis by replacing \(4\)\,MB segments of the model parameters.
In total, \(352\) unique fault-injected models were evaluated.
An overview of the results across the evaluated architectures and weight representations is provided in Fig.~\ref{fig:accuracies}.

Table~\ref{tab:consolidated_spatial} provides a consolidated overview of the spatial sensitivity analysis for selected ResNet and VGG architectures.
The remaining architectures were omitted because the post-attack accuracies for all attacked chunks were too low to support meaningful conclusions.
Here, chunks are defined as contiguous \(4\)\,MB segments of the model's weight tensors that were individually replaced during the fault-injection experiments to isolate local vulnerabilities.
The table summarizes the total number of chunks for each model and data-type configuration, together with the resulting Top-1 accuracy when targeting specific spatial regions: the Front (initial layers), the Middle (intermediate layers), and the Back (final layers or fully connected layers).

\subsubsection{ResNet Analysis}
The ResNet architectures (ResNet18, ResNet34, and ResNet50) exhibited extreme sensitivity to EMFI when using floating-point representations.

The following observations can be made:
\begin{itemize}
    \item \textbf{Floating-point (FP32/FP16):} In all tested cases, the Top-1 and Top-5 accuracies dropped close to the random-guess level, indicating a complete collapse of the models' predictive capability (Fig.~\ref{fig:accuracies}(a), (b)).
    \item \textbf{Integer (INT8/INT4):} Although these models suffered substantial accuracy degradation, they did not collapse to the random-guess level in all cases (Fig.~\ref{fig:accuracies}(c), (d)).
    \item \textbf{Spatial sensitivity:} We observed a ``front-heavy'' sensitivity profile for the networks listed in Table~\ref{tab:consolidated_spatial}. Faults injected into the initial convolutional layers resulted in severe accuracy degradation, whereas layers toward the end of the network (e.g., fully connected layers) exhibited greater resilience.
\end{itemize}

\begin{table}[ht]
\centering
\caption{Consolidated spatial sensitivity: targeted chunk accuracy (Top-1 \%). Front/Middle/Back denote the part of the network in which the attack occurred. Targeting chunks at the beginning of the network (Front) consistently leads to complete accuracy degradation.}
\label{tab:consolidated_spatial}
\begin{tabular}{|llcccc|}
\hline
\textbf{Model} & \textbf{Type} & \textbf{\# Chunks} & \textbf{Front} & \textbf{Middle} & \textbf{Back} \\ \hline
ResNet18       & INT8               & 3                     & 0.07\%             & 0.68\%                & 6.37\%            \\
ResNet34       & INT8               & 6                     & 0.10\%             & 0.12\%                & 53.05\%           \\
ResNet50       & INT8               & 7                     & 0.15\%             & 5.20\%                & 49.10\%           \\
ResNet50       & INT4               & 4                     & 0.12\%             & 0.17\%                & 22.19\%           \\ \hline
VGG11          & FP32               & 127                   & 0.05\%             & 0.05\%                & 0.12\%            \\
VGG11          & FP16               & 64                    & 0.05\%             & 0.05\%                & 62.84\%           \\
VGG11          & INT8               & 32                    & 0.15\%             & 68.14\%               & 54.13\%           \\
VGG11          & INT4               & 16                    & 0.07\%             & 65.03\%               & 20.21\%           \\ \hline
\end{tabular}
\end{table}

\subsubsection{VGG Analysis}
The VGG architecture exhibited failure modes distinct from those of ResNet.
While the FP32 models collapsed to near-random-guess accuracy, the FP16 VGG variants showed localized resilience; for example, replacing the final \(1.4\)\,MB chunk resulted in a Top-1 accuracy of approximately \(63\%\) (Table~\ref{tab:consolidated_spatial}).

Furthermore, the INT8- and INT4-quantized VGG models displayed a bimodal sensitivity profile: attacking the first and last layers caused substantial accuracy degradation, whereas attacking the middle layers preserved accuracy at the vanilla level.
The accuracy loss was therefore primarily localized to the replacement of the first and last parameter chunks, while the intermediate layers remained largely unaffected and maintained performance near the pre-attack baseline.

To summarize, the results suggest that INT8 quantization offers the best balance between inference accuracy and EMFI resilience for edge deployments. 
Beyond reducing memory footprint, the bounded nature of integer arithmetic provides an inherent form of ``numerical masking'' against the high-magnitude bit flips typical of EMFI attacks. 
The underlying mechanisms are examined further in Section~\ref{sec:discussion}.

\section{Discussion}
\label{sec:discussion}
\textbf{Number representations and fault tolerance.}
As discussed in the previous section, the difference in fault resilience between floating-point and integer representations is not merely a matter of bit-width, but is fundamentally rooted in the dynamic range and non-linear mapping of the IEEE~754 floating-point standard, in contrast to the constant-resolution scaling used in integer representations.
In addition, a phenomenon of activation-function saturation arises in floating-point models:
\begin{itemize}
    \item In FP32, a post-attack weight of $10^{38}$ produces an activation value so large that it either causes numerical overflow in the next layer or permanently ``saturates'' the neuron.
    \item In quantized networks, weights and activations are typically clipped or normalized. Even if a weight is flipped to its maximum integer value, the scaling factors used in INT8/4 quantization act as a natural shock absorber, preventing a single bit flip from dominating the entire feature map.
\end{itemize}

\textbf{Embedded memory limit.}
One potential limitation of our experimental platform is its \(4\)\,MB memory capacity, which is modest compared with that of contemporary high-performance AI accelerators.
However, this choice was deliberate and aligned with the objectives of the present study.
Our goal is not to evaluate absolute fault rates in highly optimized systems, but rather to characterize and compare the intrinsic sensitivity of different numerical representations used for neural network weights.

Highly integrated SoCs typically employ memory interleaving, redundancy, and hardware-level error-correction mechanisms, which introduce confounding factors that can obscure the direct relationship between injected faults and their algorithmic impact.
By using a linear, non-interleaved memory architecture without error-correction codes, we are able to isolate the mathematical propagation of memory faults within the network and to study representation-dependent effects in a controlled and reproducible manner.

Moreover, this memory scale is representative of the emerging Micro-AI and edge-inference domain, in which resource-constrained microcontrollers deploy neural networks in safety- and security-critical applications.
In this context, understanding the fundamental resilience of numerical representations is particularly relevant.

Extending this analysis to large-scale accelerators with complex memory hierarchies is an important direction for future work; however, such platforms would primarily affect the fault-injection surface rather than the representation-level vulnerability analyzed in this paper.

\begin{figure*}[htb]
    \centering
    \includegraphics[width=0.98\linewidth,trim={1cm 0 4cm 0},clip]{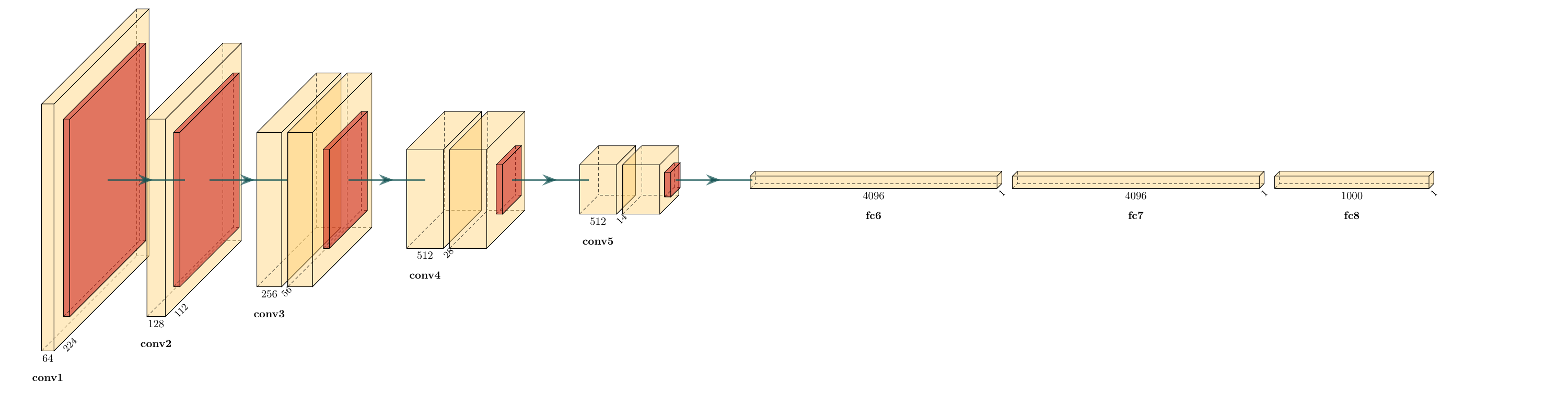}
    \caption{VGG-11 architecture, illustrating the contrast between the shapes of the convolutional and fully connected layers.}
    \label{fig:vgg11}
\end{figure*}

\textbf{VGG-11 middle-layer resilience.}
The bimodal sensitivity of VGG-11 INT8, in which attacking the middle chunks leaves the accuracy near the baseline whereas attacking the first or last chunks causes substantial degradation, can be explained by the geometry of the large fully connected layer FC6 (shape $4096 \times 25088$), which occupies approximately $98$\,MB in INT8 and approximately $49$\,MB in INT4, as illustrated in Fig.~\ref{fig:vgg11}.
A single \(4\)\,MB fault window covers approximately $4/98 \approx 4\%$ of the bytes of FC6; at the observed byte-corruption rate of approximately $15\%$, the effective fraction of FC6 weights modified per injection is only approximately $0.6\%$.
Given the high parametric redundancy of a ${\sim}100$\,M-parameter layer, this localized perturbation is insufficient to meaningfully shift the network's output distribution.
In contrast, the early convolutional layers (conv1--conv3) are both parameter-sparse and computationally critical: a \(4\)\,MB fault window covers their full weight tensors, and corruption of even a small fraction of first-layer filters destroys the intermediate feature representations on which all subsequent layers depend.

\textbf{Countermeasures.}
Countermeasures against fault injection attacks are a well-studied topic in cryptography~\cite{baksi2022survey}.
Some of these methods can be directly applied to protect embedded neural networks, such as detection circuits~\cite{zussa2014efficiency,breier2017electromagnetic} and error-correcting or error-detecting codes~\cite{guo2013recomputing,breier2019evaluating,breier2020countermeasure}.
In addition, several methods have been developed specifically to protect deep learning systems; we summarize the most prominent approaches below.

DeepDyve~\cite{li2020deepdyve} uses small pre-trained neural networks to verify the outputs of the main model.
RADAR~\cite{li2021radar} stores a 2-bit checksum for each group of weights in memory during deployment and checks the integrity of the weights at runtime.
For ResNet-18, it achieves a detection rate of $96.1\%$ with approximately $1\%$ computational overhead and $5.6$\,KB of memory overhead.
HASHTAG~\cite{javaheripi2021hashtag} also focuses on detection, but uses cryptographic hashing instead of simple checksums.
ALERT~\cite{wei2024alert} monitors and detects abnormal neuron activation patterns caused by faults.
When such anomalies are detected, ALERT activates a recovery mechanism to mitigate the impact of the attack.

\section{Conclusion}
\label{sec:conclusion}
In this work, we presented a comprehensive empirical analysis of EMFI on embedded neural networks, investigating how different numerical weight representations influence model resilience against such attacks.

Beyond measuring post-attack accuracy, we characterized the injected fault pattern itself, showing that the bit error rate is largely format-independent across all tested configurations. 
This finding establishes that the pronounced differences in resilience stem from how each representation responds to faults rather than from differences in fault density. 
In particular, the asymmetry between the \texttt{0xFE}/\texttt{0xFF} byte fractions received by floating-point and integer formats, where integer models absorb a higher proportion of maximal-value bytes yet suffer far less accuracy degradation, directly exposes the role of the IEEE 754 exponent encoding as the primary failure mechanism.

We showed that the numerical representation of weights is consequently a significant factor in model resilience: floating-point models suffer from catastrophic accuracy degradation due to NaN propagation and exponent bit flips, whereas quantized models provide an inherent form of protection through numerical clipping. 
Among the formats studied, INT8 offers the most favorable trade-off between inference accuracy, memory footprint, and EMFI resilience for edge deployments.

As a future direction, it would be worthwhile to investigate training-phase approaches for improving fault resistance. 
Although INT8 provides passive protection, new quantization-aware training schemes could be developed to specifically minimize the bit-error sensitivity of the most influential weight bits. 
Extending the analysis to in-situ inference on a microcontroller would also allow the representation-level findings reported here to be validated under realistic deployment conditions.



\vspace{0.5cm}
\noindent
\subsubsection*{Acknowledgement}
OpenAI ChatGPT, Google Gemini, and Anthropic Claude LLMs were used to improve clarity and readability of some portions of this manuscript.

\bibliographystyle{IEEEtran}
\bibliography{bibl}

\end{document}